\documentclass{aip-cp}

\pdfoutput=1
\usepackage{graphicx}
\usepackage[numbers,elide]{natbib}
\usepackage{fancybox}
\usepackage{shortvrb}
\usepackage[greek,english]{babel}
\usepackage[bf]{caption}
\addto\captionsenglish{\renewcommand{\figurename}{FIGURE}}
\MakeShortVerb\|

\usepackage{listings}
\usepackage[usenames]{color}
\definecolor{codecolor}{gray}{.9}
\definecolor{rlcolor}{cmyk}{0,1,0,0}
\begin{document}

\title{The Micromegas Project for the ATLAS New Small Wheel}

\author[aff1,aff3]{I. Manthos\corref{cor1}}
\author[aff1,aff3]{I. Maniatis}
\author[aff1,aff3]{I. Maznas}
\author[aff1,aff3]{M. Tsopoulou}
\author[aff1,aff3]{P. Paschalias}
\author[aff1,aff3]{\mbox{T. Koutsosimos}}
\author[aff2,aff3]{S. Kompogiannis}
\author[aff1,aff3]{Ch. Petridou}
\author[aff1,aff3]{S.E. Tzamarias}
\author[aff1,aff3]{\mbox{K. Kordas}}
\author[aff1,aff3]{\mbox{Ch. Lampoudis}}
\author[aff2,aff3]{I. Tsiafis}
\author[aff1,aff3]{D. Sampsonidis}

\affil[aff1]{School of Physics, Aristotle University of Thessaloniki, University Campus, GR-54124, Thessaloniki, Greece}
\affil[aff2]{School of Mechanical Engineering, Aristotle University of Thessaloniki, GR-54124, Thessaloniki, Greece}
\affil[aff3]{Center for Interdisciplinary Research and Innovation (CIRI-AUTH), Thessaloniki, GR-57001,Greece}

\corresp[cor1]{Corresponding author: ioannis.manthos@cern.ch}
\maketitle

\begin{abstract}
The MicroMegas technology was selected by the ATLAS experiment at CERN to be adopted for the Small Wheel upgrade of the Muon Spectrometer, dedicated to precision tracking, in order to meet the requirements of the upcoming luminosity upgrade of the Large Hadron Collider. A large surface of the forward regions of the Muon Spectrometer will be equipped with 8 layers of MicroMegas modules forming a total active area of 1200\,m\textsuperscript{2}. The New Small Wheel is scheduled to be installed in the forward region of $1.3<\vert \eta \vert <2.7$ of the ATLAS detector during the second long shutdown of the Large Hadron Collider. The New Small Wheel will have to operate in a high background radiation environment, while reconstructing muon tracks as well as furnishing information for the Level-1 trigger. The project requires fully efficient MicroMegas chambers with spatial resolution down to $100\,\textrm{\selectlanguage{greek}m\selectlanguage{english}m}$, a rate capability up to about 15\,kHz/cm\textsuperscript{2} and operation in a moderate (highly inhomogeneous) magnetic field up to B=0.3\,T. The required tracking is linked to the intrinsic spatial resolution in combination with the demanding mechanical accuracy. An overview of the design, construction and assembly procedures of the MicroMegas modules will be reported. 
\end{abstract}
\section{INTRODUCTION}
The Large Hadron Collider (LHC) \citep{lhc1, lhc} will be upgraded in several stages (Phase-1, Phase-2). After the second long shutdown (LS2) in 2019-2020 the luminosity of the accelerator will be increased up to $2 - 3 \times 10^{34}\,\mathrm{cm^{-2} s^{-1}}$, which will allow ATLAS detector \citep{atlas} to collect up to 100\,fb\textsuperscript{-1}/year data. In order to take advantage those record parameters of the LHC efficiently, the innermost stations of the forward region of  the ATLAS muon spectrometer (Small Wheel - SW) will be replaced within the same time schedule of the LHC upgrade by a New Small Wheel (NSW) \citep{nsw} during the Phase-1 upgrade. The muon spectrometer has to retain the ability to reconstruct particle tracks with high precision as well as to provide information to the Level-1 (L1) trigger. In particular, the offline track reconstruction requires a spatial resolution of about $100\,\textrm{\selectlanguage{greek}m\selectlanguage{english}m}$ per detector layer and an angular resolution of approximately 1\,mrad for the online track segment reconstruction for the L1 trigger at background rates up to 15\,kHz/cm\textsuperscript{2}. For this purpose a new set of detectors have been designed and are being constructed for the NSW, implementing two technologies, one primarily devoted to the L1 trigger function (small-strip Thin Gap Chambers, sTGC \citep{stgc}) and one dedicated to precision tracking (MicroMegas detectors - MM \citep{mm}). The MM detectors have outstanding precision tracking capabilities due to their small gap (5\,mm) and strip pitch ($450\,\textrm{\selectlanguage{greek}m\selectlanguage{english}m}$). 
\section{THE MICROMEGAS DETECTORS FOR THE NSW}
\begin{figure}
\centering
\includegraphics[width=0.9\textwidth]{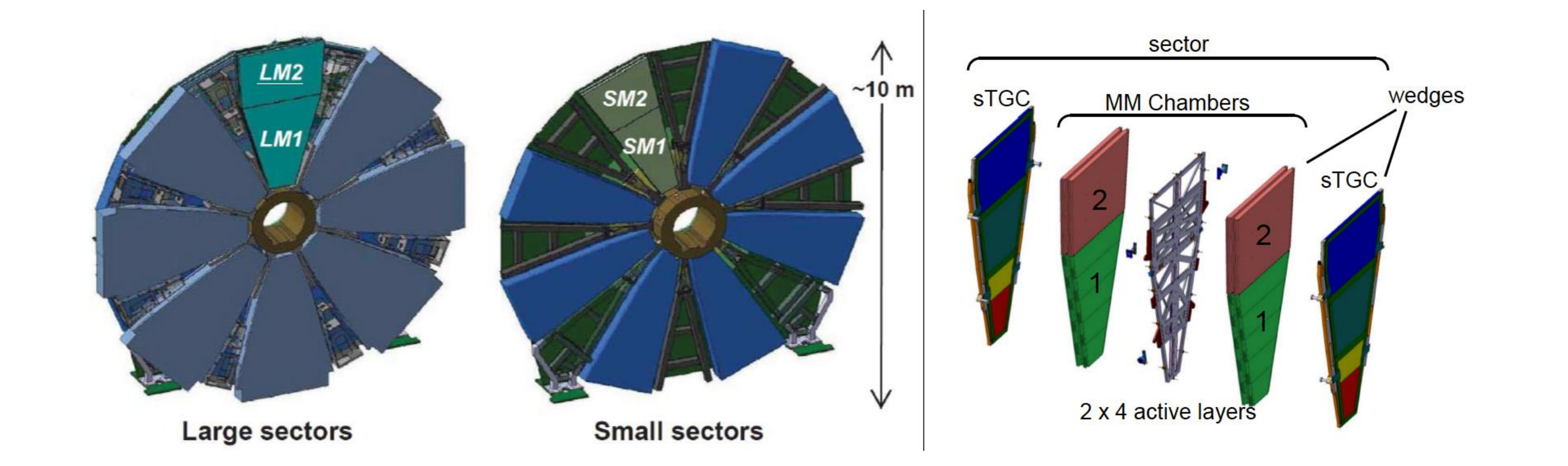}
\caption{Overall view of the New Small Wheel. Left: Views of the wheel, highlighting the modules large sectors LM1 and LM2 and the modules of the small ones SM1 and SM2. Right: Exploded view of a Large sector, with two sTGC and two MM quadruplets on both sides of the spacer frame.}
\label{fig:fig1}
\end{figure}

\subsubsection{The Design of the New Small Wheel} 

\begin{figure}
\centering
\begin{minipage}{.55\textwidth}
\centering
\includegraphics[width=1.\textwidth]{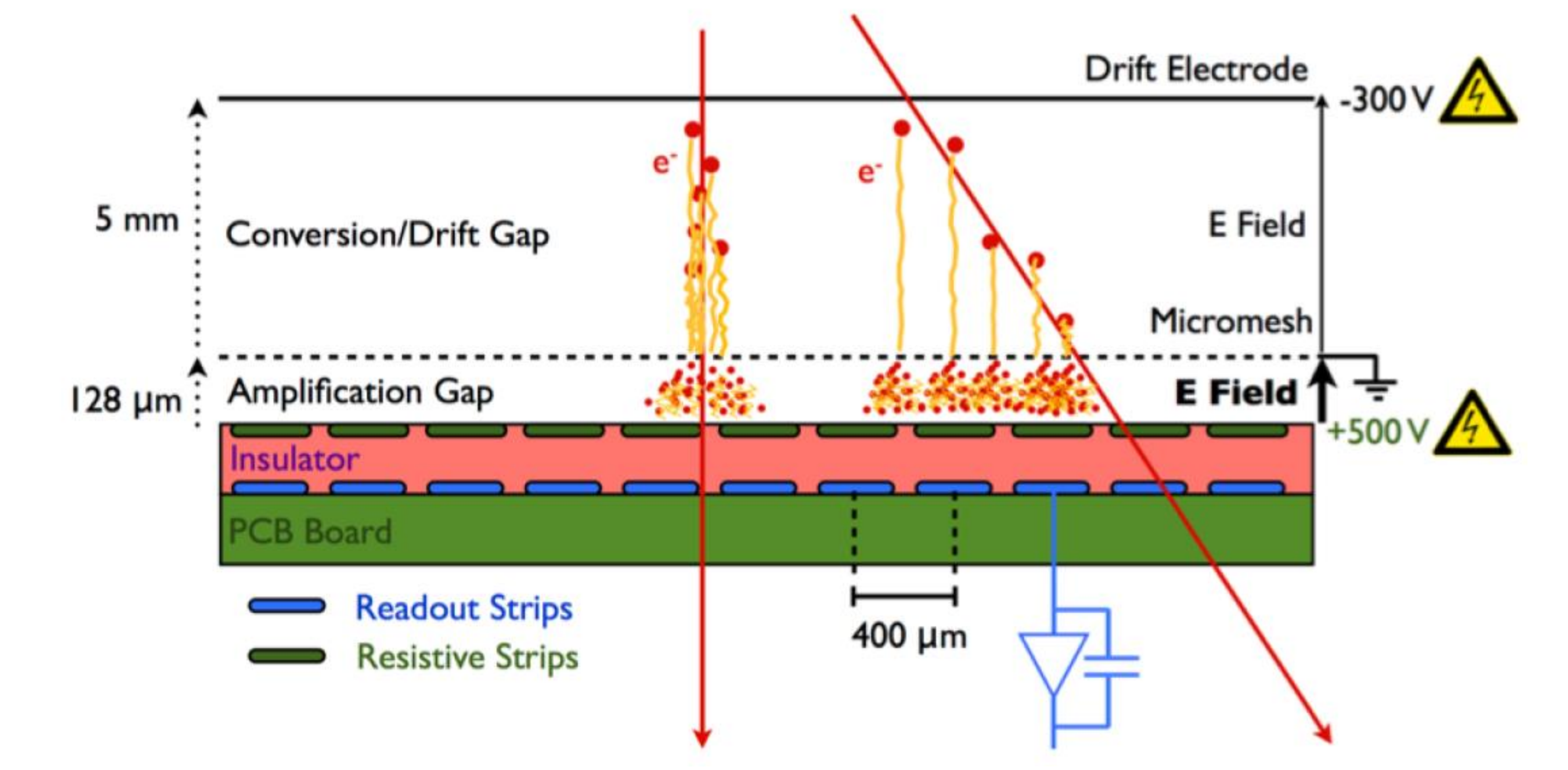}
\end{minipage}
\begin{minipage}{.45\textwidth}
\centering
\includegraphics[width=.75\textwidth]{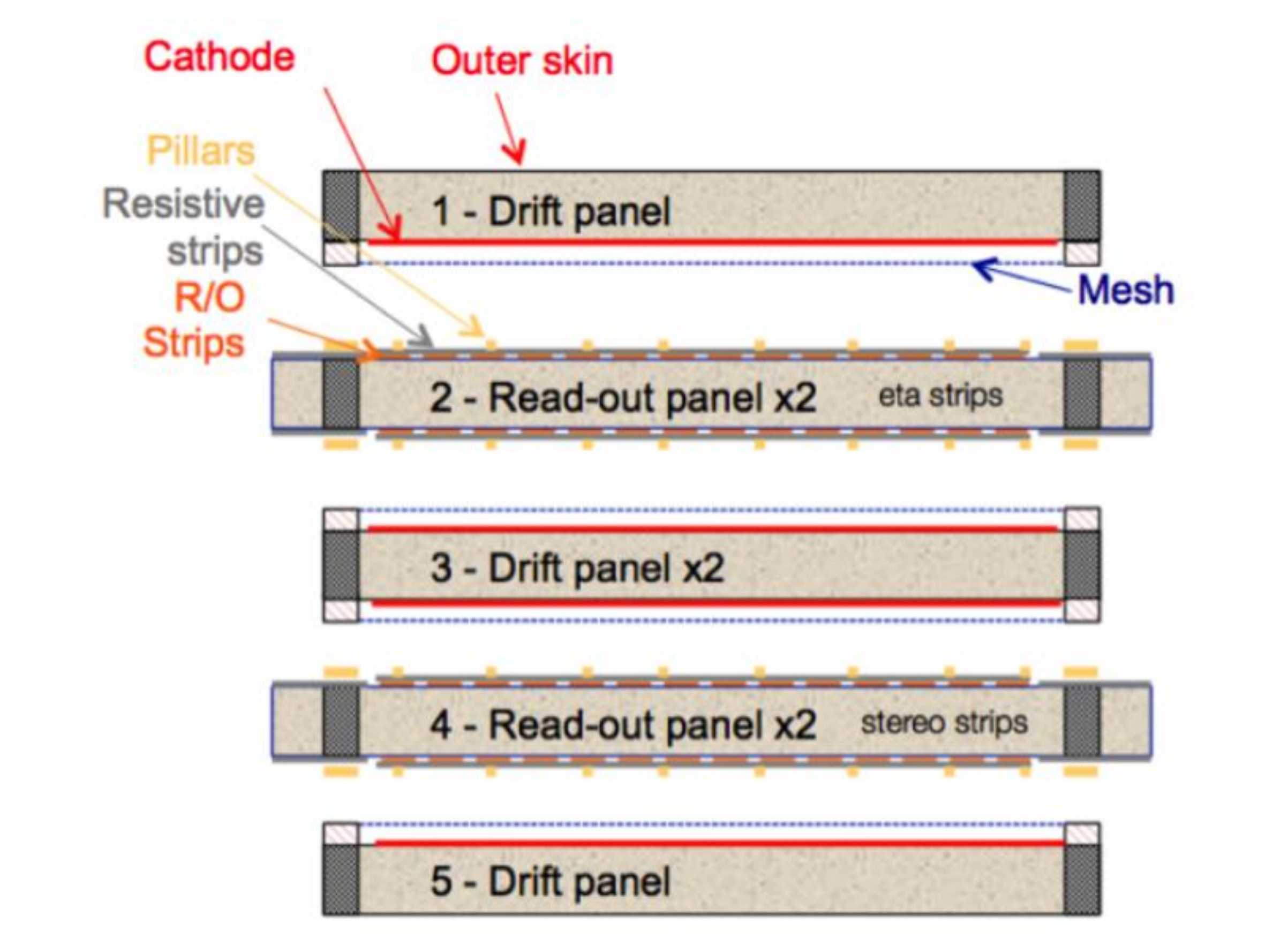}
\end{minipage}
\caption{Left: A scheme of a single MM layer. Right: A schematic view of the five panels of a MM forming a quadruplet.}
\label{fig:fig2}
\end{figure}

The NSW of  the ATLAS muon spectrometer will be comprised of two wheels equipped with sTGC and MM chambers. The wheel has eight large and eight small sectors (wedges) partially overlapping and fixed on a metallic circular structure, as shown in Fig. \ref{fig:fig1} (left). The sector consists of two MM wedges attached on both sides of the metallic structure and sandwiched by two sTGC wedges as shown in Fig. \ref{fig:fig1} (right). MM and sTGC chambers in the wedges have four active detector layers forming quadruplets. Therefore the NSW has eight layers of MM and eight layers of sTGC detectors. Each MM wedge is segmented in two parts of different size trapezoids, each covering a different region in $\eta$. This results in four types of MM chambers: SM1 and SM2 as Small Sectors Modules, LM1 and LM2 as the Large Sectors Modules, corresponding to chamber sizes of $\sim 2\,\mathrm{m}^{2}$ and $\sim 3\,\mathrm{m}^{2}$ respectively. The construction of the four different MM modules has been shared between four laboratory consortia, one for each type of chambers: INFN, Italy for the SM1, Germany for the SM2, Saclay, France for the LM1 and  Dubna, Russia, Thessaloniki, Greece and  CERN for the LM2. In this paper we present the construction of the LM2 modules as well as the quality control procedure followed during the construction. 
\subsubsection{The MicroMegas Detector for the NSW}
A single MicroMegas is a position sensitive gaseous detector with planar parallel electrodes and consists of three planes, the cathode, the anode and the micromesh as shown in Fig. \ref{fig:fig2} (left). The mesh divides the gas volume between anode (readout) and cathode (drift) in two areas, the conversion/drift gap of 5\,mm where the incident muon ionizes the gas molecules and creates primary electrons and the amplification gap of $128\,\textrm{\selectlanguage{greek}m\selectlanguage{english}m}$ where electron avalanches occur. The anode plane is based on printed circuit boards (PCB), with photo-lithographically etched copper strips and a layer of resistive strips on a kapton foil glued on the copper strips for the discharge protection. The resistive strips have a resistivity of 10 to $20\,\mathrm{M\Omega/cm}$. The readout strips have a pitch of $450\,\textrm{\selectlanguage{greek}m\selectlanguage{english}m}$. The mesh is supported by $128\,\textrm{\selectlanguage{greek}m\selectlanguage{english}m}$ high pillars, which guarantee the uniformity of the amplification gap. The cathode is also a PCB having a copper surface.
\subsubsection{The MicroMegas Chamber}
A MM chamber consists of four gas gaps forming a quadruplet of MicroMegas detectors. The MM chambers need to fulfil mechanical requirements in order to be able to reconstruct a muon momentum with a resolution of 15\% at 1\,TeV in ATLAS. This implies excellent quality of the materials used for the construction, strict quality control of the PCBs, as well as special construction methods. The precision for the strip position in $\eta$ (precision coordinate) should be 30 microns r.m.s., and 80 microns r.m.s. for the strip position in Z (perpendicular to the detection plane). Fig. \ref{fig:fig2} (right) shows a schematic view of a quadruplet. Five panels, providing the required stiffness, bound the four active MM gaps. The panels are trapezoidal in shape. Each quadruplet has one ``$\eta$'' readout panel (two detection planes in $\eta$ direction) having the strips parallel to the trapezoid bases in order to measure $\eta$ coordinate and one ``stereo'' panel (two stereo detection planes, where strips have an angle of  $\pm 1.5^\circ$) to measure $\eta$ coordinates but also reconstruct the second coordinate $\phi$ with few mm resolution. There are two drift panels having one side covered with copper implementing the cathodes for the outer gaps and one central drift panel having both sides covered with copper. The mesh is glued on the drift panels via an appropriate mesh frame and attached to the pillars when the quadruplet is assembled.  
\subsubsection{The MicroMegas Panels}
\begin{figure}
\centering
\includegraphics[width=0.7\textwidth]{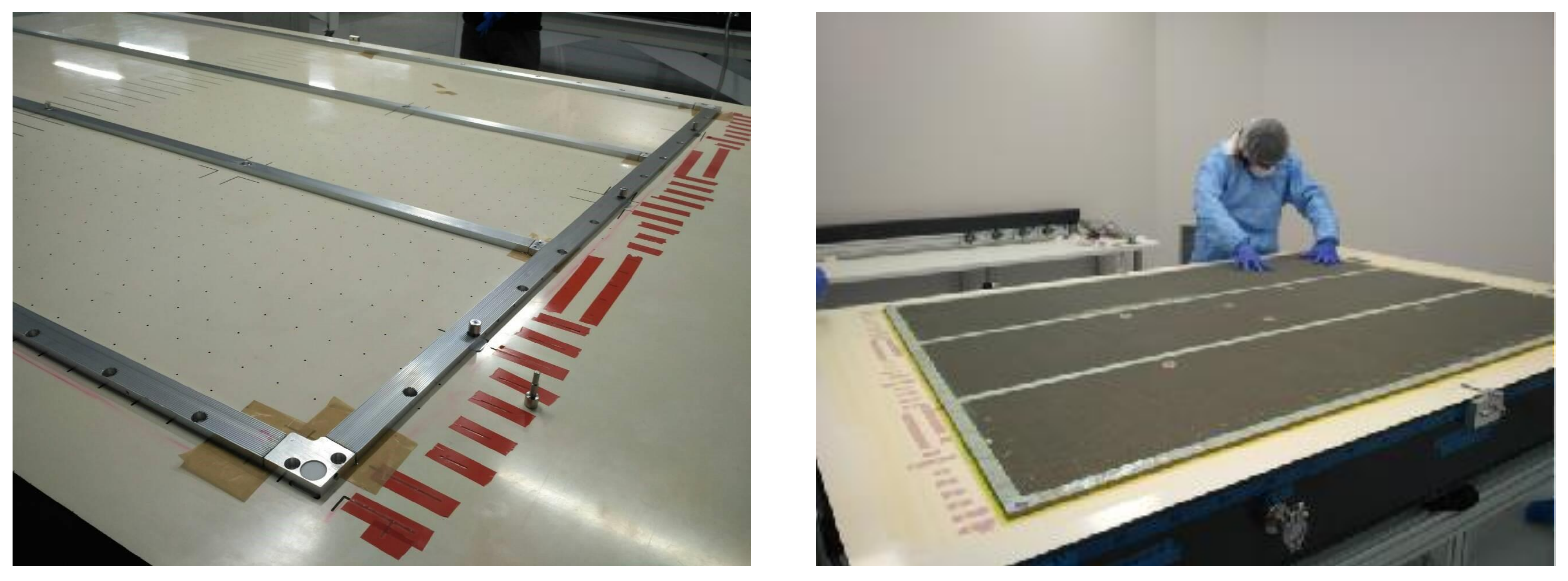}
\caption{Left: The aluminum frame of the drift panel. Right: The positioning of the honeycomb during the drift panel construction.}
\label{fig:fig4}
\end{figure}

A panel is a stiff light structure consisting of  an aluminum frame, aluminum honeycomb 10.1\,mm thick with 6\,mm hexagonal cells and having on the two outer surfaces 0.5\,mm FR4 material of the PCBs. The frame consists of aluminum bars with 10.0\,mm thickness, perimetric and cross as reinforcement bars with four special aluminum corners which implement the gas inlet and outlet to the MM gap. All the aluminum parts are tested to fulfil the mechanical requirements for the thickness and bending deformations. The frame is prepared in advance by gluing all the parts on a precise assembly table using special dowel pins to align the parts properly (Fig. \ref{fig:fig4}, left). 
\subsubsection{Drift Panel Construction}
\begin{figure}
\centering
\includegraphics[width=0.7\textwidth]{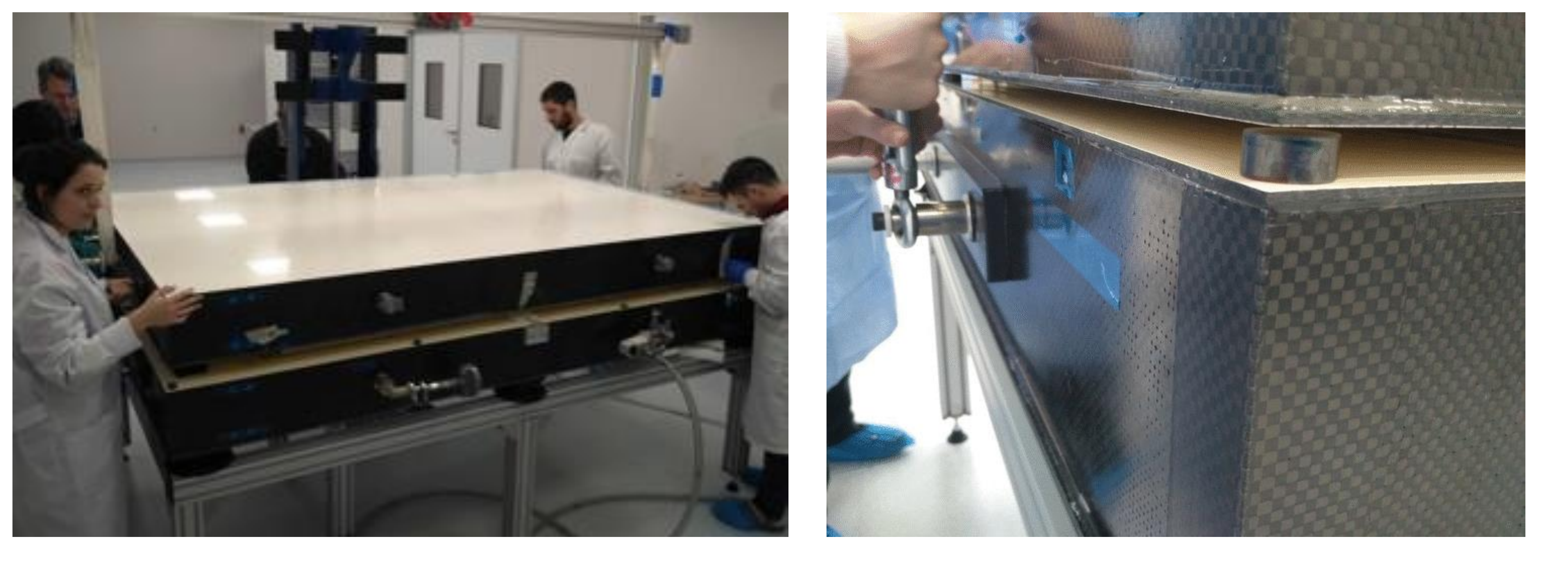}
\caption{Left: Table II on top of table I during drift panel construction. Right: The precision spacers between the two tables.}
\label{fig:fig5}
\end{figure}

According to the described structure of the quadruplet, two external and one central drift panels are produced for each LM2 Module (quadruplet) in a class D clean room, at the University of Thessaloniki. The nominal tolerances for the panel planarity are $37\,\textrm{\selectlanguage{greek}m\selectlanguage{english}m}$ in r.m.s., equivalent to $\pm 110\,\textrm{\selectlanguage{greek}m\selectlanguage{english}m}$ mechanical tolerance. The panel construction is done using the vacuum table method in one step. The PCBs are indexed on the vacuum tables via precision holes on the tables, using 5 mm dowel pins and consequently sucked at an under-pressure of 100 - 150\,mbar in order to take advantage of the planarity of the tables ($\pm 30\,\textrm{\selectlanguage{greek}m\selectlanguage{english}m}$). The glue is distributed on the PCBs as well as on the aluminum frame. The frame is positioned on the PCBs standing on the vacuum table I and indexed using pins on the four corners. After positioning the honeycomb pieces (Fig. \ref{fig:fig4}, right), on the table I, the table II is placed on top of the first one standing on ten high precision spacers which define the panel thickness, as shown in Fig. \ref{fig:fig5}. The under-pressure is maintained for $\sim 20\,\mathrm{h}$ during glue curing. Next day, the upper table is removed and the panel is completed.
\subsubsection{Drift Panel Completion}
The finalization started with some minor fixing of the panels, like removing the PCB in excess and the remaining glue as well. The mesh frames are then glued on the drift plane with the required accuracy of $\pm 200\,\textrm{\selectlanguage{greek}m\selectlanguage{english}m}$ in the plane and $\pm 25\,\textrm{\selectlanguage{greek}m\selectlanguage{english}m}$ in height, guaranteed both by the precision of the mesh frame profile and by plastic rulers referring to the closing holes. In order to have the mesh frame profile connected to the panel grounding, screws are also inserted and establish the contact to the panel aluminum frame. Next the interconnection spacers are glued with $\pm 25\,\textrm{\selectlanguage{greek}m\selectlanguage{english}m}$ precision in height with a special tool. Following, the gas distribution pipes are mounted and glued in contact with the inner side of the mesh frame long and short sides. Finally, the HV connectors are glued in the outer side of the aluminum frame of the panel and soldered in the copper of the PCB.
\section{QUALITY ASSURANCE OF NSW MICROMEGAS DRIFT PANELS}
\subsubsection{Planarity and Thickness}
\begin{figure}
\centering
\includegraphics[width=0.9\textwidth]{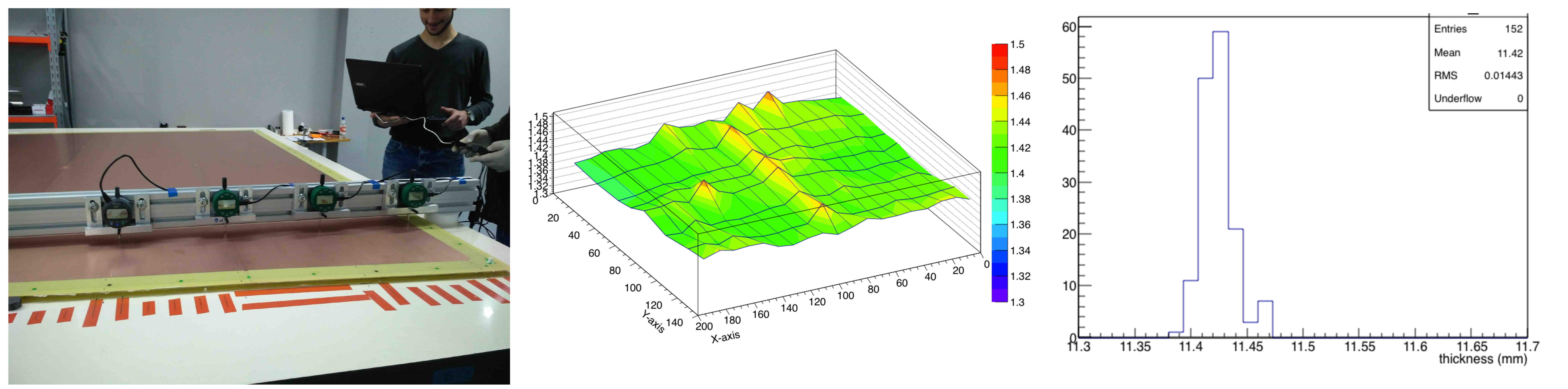}
\caption{Left: Planarity and thickness measurement with the limbo tool. Center: Contour plot of the thickness measurements. Right: Distribution of the panel thickness.}
\label{fig:fig6}
\end{figure}

All panels are measured for their planarity and thickness. For these measurements a limbo tool has been developed, which consists of a rigid Al profile, instrumented with four height gauges, read out by a PC (Fig. \ref{fig:fig6}, left). The measurement of the thickness is performed on the vacuum table with the vacuum ON, having an under pressure of about 100\,mbar, while the measurement for the planarity is performed without vacuum. The gauges are calibrated on a granite bar before every measurement and then the panel surface is scanned taking measurements at a grid of 152 points in total. Data are recorded and analyzed showing the planarity and the thickness of a panel. Fig. \ref{fig:fig6}, right, shows the planarity results of cathode (copper) side of a panel. The r.m.s. is $14\,\textrm{\selectlanguage{greek}m\selectlanguage{english}m}$ well within the specifications. 
\subsubsection{Gas Tightness Certification }
\begin{figure}
\centering
\includegraphics[width=0.9\textwidth]{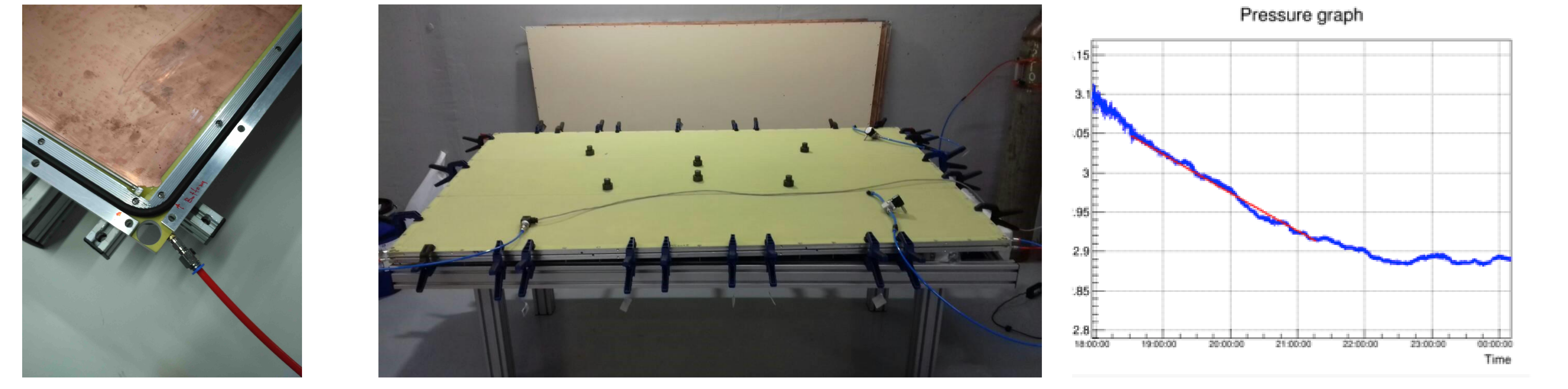}
\caption{Left: The o-ring, placed between the mesh frame and the gas gap aluminum frame. Center: The gas tightness setup. Right: The pressure drop is monitored and fitted.}
\label{fig:fig7}
\end{figure}

Before the mesh gluing on the drift panel, the gas leakage test is performed, to ensure the tightness of the panel. For this test, a pair of gas-tight dummy panels is used which serves as a vessel for the drift panel under test. In the case of an external drift panel, one dummy panel is used, as it participates to one MM detector in the quadruplet and only one active area has to be checked, while in the central panel case both sides have to be checked. The sequence of the gas tightness test for an external drift panel is the following: a) the external panel is placed on the dedicated tool for the panel holding and is extensively cleaned, b) a 7\,mm diameter o-ring is positioned on the periphery of the panel, c) the dummy panel is positioned in such a way to close the drift panel, creating the gas gap, d) a gas gap aluminum frame, 5\,mm thick, outside the o-ring is positioned and the o-ring is suppressed and seals the gas gap between the two panels (Fig. \ref{fig:fig7}, left), e) the two panels are clamped together using both clamps and screws, f) six dedicated interconnection caps equipped with o-ring are screwed on both sides of the interconnection holes tightening the interconnection areas of the panel and reducing the deformation of the panel when the gas gap is filled with Argon. As the setup is filled with Ar at about 4\,mbar overpressure (the working pressure on NSW will be 3\,mbar), the pressure drop is monitored typically for 24\,h. The data are corrected for temperature variation and for the ambient pressure and an exponential fit is performed to the pressure drop rate. The requirement for the gas leakage of the NSW panels to be certified is set to 0.6\,mbar/h at pressure of 3\,mbar, mainly in order to avoid contamination of the MM chamber with oxygen that would affect the ionization rate. Fig. \ref{fig:fig7} shows the gas leak test setup (center) and the measurement of the pressure drop for a period of six hours and the exponential fit for the data range around 3\,mbar pressure (right). 
\section{MESH ASSEMBLY}
\begin{figure}
\centering
\includegraphics[width=0.85\textwidth]{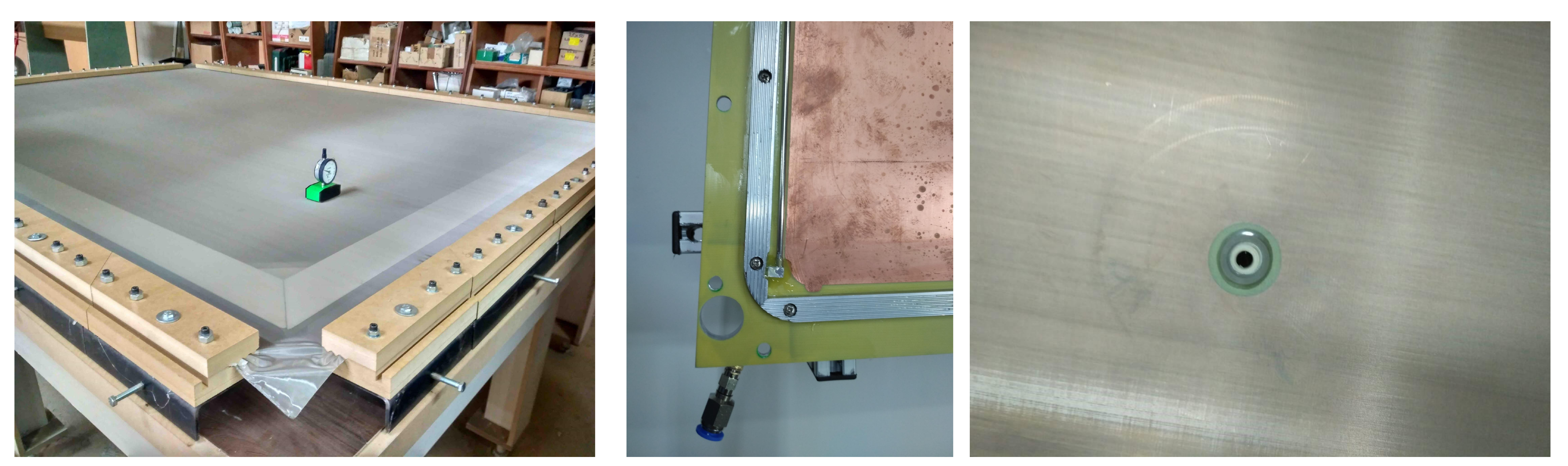}
\caption{The mesh stretching table (left), the mesh frame glued in the panel (center) and the perforated mesh in the interconnection area (right).}
\label{fig:fig8}
\end{figure}

\begin{figure}
\centering
\includegraphics[width=0.8\textwidth]{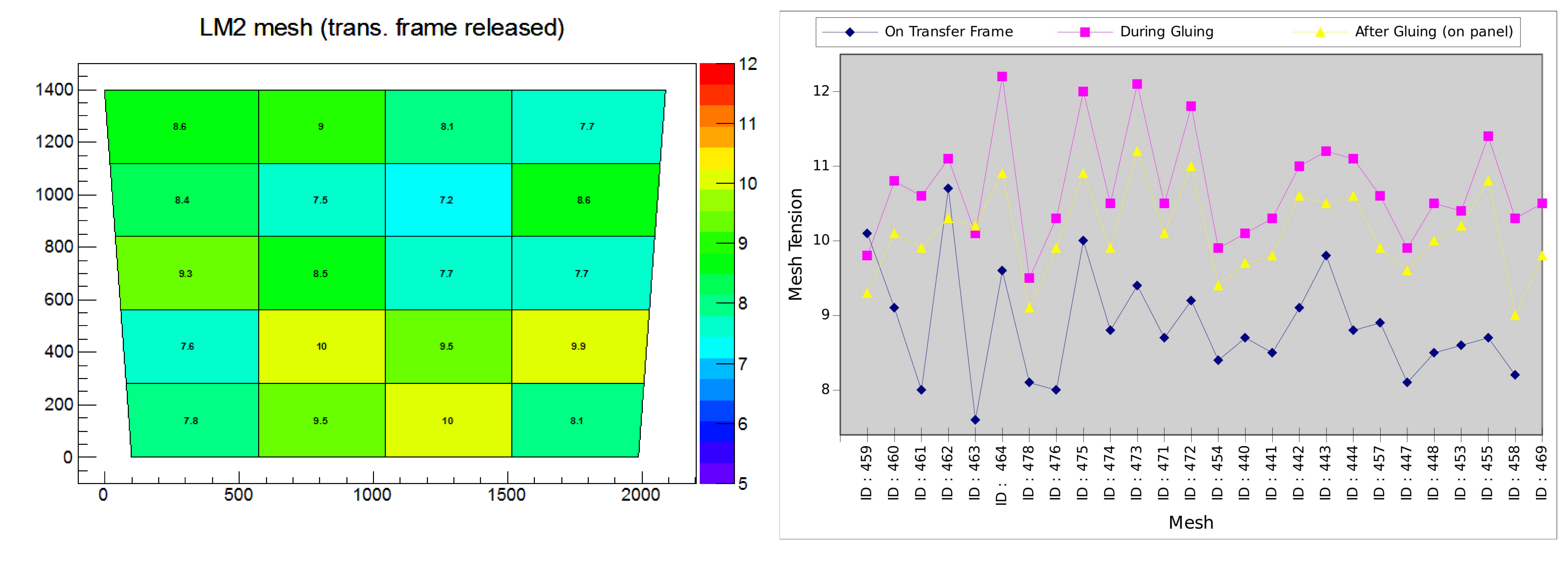}
\caption{Mesh mapping (left) and the measured average tension of the first 28 meshes (right).}
\label{fig:fig9}
\end{figure}

The electrical transparency of the stainless steel micro-mesh, necessary for the passage of drift electrons, depends on both its mechanical structure as well as on the ratio between the amplification and drift electric fields. In order to ensure the homogeneity of the amplification gaps and avoid sags between the pillars, the mesh must be precisely tensioned and glued on the drift panel. The positioning of the mesh at the right distance from the anode is guaranteed by the stretch of the mesh along the edges and by the electrostatic forces due to the large electric field between mesh and anode that pulls the mesh to the pillars. This method, called floating mesh, is a novel technique, since all previous MMs of smaller dimensions had been built with the so called bulk technology. In the floating mesh concept, the quadruplet can be reopened since the mesh is not glued to the read-out PCB. The mesh used in the NSW project is made of $28\,\textrm{\selectlanguage{greek}m\selectlanguage{english}m}$ diameter wires, woven with 325 lines per inch (corresponding to a pitch of $78\,\textrm{\selectlanguage{greek}m\selectlanguage{english}m}$). During the construction, the mesh is stretched to the desired tension (Fig. \ref{fig:fig8}, left), then six holes are made at the points where the interconnections must pass through (Fig. \ref{fig:fig8}, right), and finally it is glued on the aluminum frames (the mesh frames) on the drift panels (Fig. \ref{fig:fig8}, center). The nominal mesh tension is in the range of 7 - 10\,N/cm with a uniformity of $\pm 10\%$ after gluing on the drift panel. For the mesh stretching, a stretching table ($\sim 2 \times 2.7\,\mathrm{m}^{2}$) was built. The table is equipped with a total of 24 clamps, each 37\,cm long, placed along the four sides. Clamps are holding the mesh through screwing nuts and are equipped with load cells that can be manually pulled through screwing nuts as well, to apply tension to the mesh. In order to move the mesh from the stretching table to the panel, while keeping it at the achieved tension, reusable transfer frames were built, where the mesh is glued after the stretching to the nominal tension. Due to a decrease of the tension on the mesh after cutting the mesh, in order to release from the clamps and a further decrease when cutting the mesh on the mesh frame of the panel after gluing, resulting to $\sim 10 - 15\%$, the initial tension on the stretch table is adjusted to $\sim 9.5 - 10\,\mathrm{N} \cdot \mathrm{m}$. A full map of the tension is produced by measuring the applied tension with an analog gauge after gluing the mesh on transfer frame, during and after gluing the mesh on the mesh frame of the panel (Fig. \ref{fig:fig9}, left).
The final step in the drift panel completion is the transfer and gluing of the pre-stretched mesh from the transfer frame onto the panel mesh frame. The glue is deposited on the inclined surface of the mesh frame with the help of a dedicated tool. Once the glue is distributed all along the mesh frame, the stretched mesh on its transfer frame is lowered onto the drift panel. A last inspection of the mesh tension is carried out and if low tension is observed, extra weight is placed at the perimeter of this specific area of the panel. After glue curing overnight, the mesh is cut with a sharp scalpel, and the final product is ready for the final certification measurements. The tension values and the uniformity of the glued meshes on the drift panels are presented in Fig. \ref{fig:fig9} (right).

\section{FINAL TESTS, STORAGE AND TRANSPORTATION}
The final test that has to be performed on the drift panel is the electrical measurements in order to assure the electric insulation between the mesh and the copper of the PCB (cathode electrode). When applying 500\,V the current measured has to be less than 10\,nA. After this test, a final thorough inspection is performed. With the completion of the construction and certification of the drift panels, they are stored in dedicated prepared storage boxes, until their transportation to the JINR in Dubna, Russia. There, the read-out panels are constructed and the 5 panels of each Module are assembled in a quadruplet. The completed quadruplet is then transported to CERN for the final tests and the installation to the NSW.

\section{CONCLUSIONS}
The Micromegas detectors will be used in the NSW upgrade of the ATLAS Muon Spectrometer enabling to retain its excellent performance after the next LHC upgrade. The construction method as well as the Quality Assurance of the drift panels of the LM2 chambers which are produced in the Aristotle University of Thessaloniki, Greece, has been presented. All panels produced so far found within the ATLAS requirements. 
\section{ACKNOWLEDGEMENTS}
This work was supported by the European Union and the ESPA 2014-2020 National Fund for Research Infrastructures, Grant No 5029538 from the Structural Funds, European Regional Development Funds (ERDF) and European Structural Funds (ESF), Greece.
\bibliographystyle{unsrt}
\bibliography{mm_bib}

\end{document}